\documentclass[11pt,conference]{IEEEtran}

\usepackage{placeins}
\usepackage{graphicx}
\usepackage[export]{adjustbox}
\usepackage{tikz}
\usepackage{tabularx}
\usepackage{multirow}
\usepackage{siunitx}
\usepackage{enumitem}
\usepackage{amssymb}
\usepackage{tabularray}
\UseTblrLibrary{varwidth}
\usepackage{fancyhdr}
\usepackage[acronym]{glossaries}
\newacronym{Mem}{Mem}{Model Size}
\newacronym{PT}{PT}{Processing Time}

\newacronym{AODV}{AODV}{Adhoc On Demand Vector}
\newacronym{RFE}{RFE}{Recursive Feature Elimination}
\newacronym{MI}{MI}{Mutual Information}
\newacronym{SC}{SC}{Smart Contract}
\newacronym{GA-SVM}{GA-SVM}{ Genetic Algorithm-based Support Vector Machine} 
\newacronym {GA-DT}{GA-DT}{Genetic Algorithm-based Decision Tree}
\newacronym {MLP}{MLP}{Multilayer Perceptron}
\newacronym{KNN}{KNN}{K-Nearest Neighbors}
\newacronym{IPFS}{IPFS}{Interplanetary File System}
\newacronym{CompNB}{CompNB}{Complement NB}
\newacronym{NC}{NC}{Nearest Centroid}
\newacronym{VBFT}{VBFT}{Verifiable Byzantine Fault Tolerance}
\newacronym{BFT}{BFT}{Byzantine Fault Tolerance}
\newacronym{OCE}{OCE}{Ontology Consensus Engine}
\newacronym{VRF}{VRF}{Verifiable Random Function}

\newacronym{TSS}{TSS}{Transmitted Signal Strength}
\newacronym{TS}{TS}{Trust Score}
\newacronym {MN}{MN}{monitor node}
\newacronym {MITM}{MITM}{Man-in-the-Middle}
\newacronym {RSS}{RSS}{Received Signal Strength}
\newacronym {PSR}{PSR}{Packet Sending Rate}
\newacronym {PFR}{PFR}{packet forwarding rate} 
\newacronym {FD}{FD}{forwarding Delay }
\newacronym {ECA}{ECA}{Energy Consumption Amount}
\newacronym  {NF}{NF}{Node Affinity}
\newacronym{GBCRP}{GBCRP}{GAN and Block Chain based
secured Routing Protocol} 
\newacronym{GBDT}{GBDT}{gradient boosting decision tree}

\newacronym{SOA}{SOA}{service-oriented architecture}
\newacronym{RNN}{RNN}{recurrent neural networks}
\newacronym{VVLC}{VVLC}{Vehicular Visible Light Communication}
\newacronym{CM}{CM}{cluster member}
\newacronym{V2X}{V2X}{vehicle to everything communications} 
\newacronym{V2V}{V2V}{vehicle to vehicle} 
\newacronym{LTE}{LTE}{Long-Term Evolution}
\newacronym{DSRC}{DSRC}{dedicated short-range communications}
\newacronym{OWC}{OWC}{optical wireless communication}
\newacronym{FSO}{FSO}{free spcae optical communication}
\newacronym{PPM}{$M$-PPM}{multilevel pulse position modulation}
\newacronym{GPS}{GPS}{Global Positioning System}
\newacronym{DC-OFDM}{DCO-OFDM}{direct-current-optical-\gls{OFDM}}
\newacronym{PAM}{$M$-PAM}{pulse amplitude modulation}
\newacronym{EPPM}{EPPM}{Expurgated pulse position modulation}
\newacronym{PAPR}{PAPR}{peak to average power ratio}
\newacronym{DPPM}{DPPM}{Differential pulse position modulation}
\newacronym{OPPM}{OPPM}{Overlapping pulse position modulation}
\newacronym{VPPM}{VPPM}{variable PPM}
\newacronym{ISI}{ISI}{inter-symbol interference}
\newacronym{MPPM}{MPPM}{Multiple pulse position modulation}
\newacronym{FDGAN}{FDGAN}{Fully Distributed Generative
Adversarial Networks}
\newacronym{GOSS}{GOSS}{Gradient-Based One-Side Sampling} 

\newacronym{EFB}{EFB}{Exclusive Feature Bundling}
\newacronym{VPFT}{VPFT}{Verifiable Byzantine Fault Tolerance} 
\newacronym{IFS}{IFS}{Interplanetary File System}
\newacronym{OSI}{OSI}{Open Systems Interconnection}
\newacronym{DoS}{DoS}{Denial of Service}
\newacronym{NFR}{NFR}{non-functional requirement}
\newacronym{MAC}{MAC}{Media Access Control}
\newacronym{IT}{IT}{Information Technology}
\newacronym{IoT}{IoT}{Internet of Things}
\newacronym{WSN}{WSN}{Wireless Sensor Network}
\newacronym{iid}{iid}{independent and identically distributed}
\newacronym{DDoS}{DDoS}{Distributed denial of service}
\newacronym{IDS}{IDS}{Intrusion Detection System}
\newacronym{BC}{BC}{Blockchain}
\newacronym{BS}{BS}{base station}
\newacronym{CH}{CH}{cluster head}
\newacronym{QoS}{QoS}{quality of service}
\newacronym{ML}{ML}{Machine Learning}
\newacronym{LR}{LR}{Logistic Regression}
\newacronym{NB}{NB}{Naive Bayes}
\newacronym{K-NN}{K-NN}{K-Nearest Neighbors}
\newacronym{SVM}{SVM}{Support Vector Machine}
\newacronym{DT}{DT}{Decision Trees}
\newacronym{ANN}{ANN}{Artificial Neural Network}
\newacronym{RF}{RF}{Random Forests}
\newacronym{DL}{DL}{Deep learning}
\newacronym{RL}{RL}{reinforcement learning}
\newacronym{DRL}{DRL}{deep reinforcement learning}
\newacronym{PoW}{PoW}{Proof-of-Work}
\newacronym{AN}{AN}{aggregating node}

\newacronym{P2P}{P2P}{peer-to-peer}

\newacronym{ECDSA}{ECDSA}{elliptic-curve digital signature algorithm} 
\newacronym{tsp}{tsp}{transactions per second}
\newacronym{CPU}{CPU}{central processing unit}

\newacronym{FL}{FL}{Federated Learning}
\newacronym{PPV}{PPV}{positive prediction value}
\newacronym{ERR}{ERR}{Error rate}

\newacronym{GM}{GM}{geometric Mean}
\newacronym{RMSE}{RMSE}{root mean square error}
\newacronym{NRMSE}{NRMSE}{normalized \gls{RMSE}}
\newacronym{ROC}{ROC}{receiver operating characteristics}
\newacronym{RAM}{RAM}{random access memory}
\newacronym{KB}{KB}{kilobytes}
\newacronym{BW}{BW}{bandwidth}
\newacronym{Acc}{Acc}{classification accuracy}
\newacronym{Pd}{$P_d$}{probability of detection}
\newacronym{Pfa}{$P_{fa}$}{probability of false alarm}

\newacronym{Pmd}{$P_{md}$}{probability of misdetection}
\newacronym{HECC}{HECC}{Hyperel-liptic Curve Cryptography}
\newacronym{SDN}{SDN}{software-defined  networking} 
\newacronym{LSTM}{LSTM}{long short-term memory}
\newacronym{PoA}{PoA}{ Proof of Authority}
\newacronym{PoS}{PoS}{Proof of Stake}
\newacronym{DPoS}{DPoS}{Delegated Proof of Stake}
\newacronym{PoC}{PoC}{Proof of Capacity}
\newacronym{PBFT}{PBFT}{practical Byzantine fault tolerance}

\newacronym{CA}{CA}{Certificate Authority}
\newacronym{AES}{AES}{Advanced Encryption Standard} \newacronym{GA}{GA}{Genetic Algorithm}

\newacronym{CNN}{CNN}{Convolutional Neural Network}
\newacronym{HMM}{HMM}{Hidden Markov Model}
\newacronym{DNN}{DNN}{Deep Neural Network}
\newacronym{GAN}{GAN}{Generative Adversarial Networks}
\newacronym{ADC}{ADC}{Analog to Digital Converter}
\newacronym{PoET}{PoET}{Proof of Elapsed Time}
\newacronym{tps}{tps}{transactions per second} 
\newacronym{CPS}{CPS}{Cyber-Physical Systems}
\newacronym{HGB}{HGB}{Histogram Gradient Boost}
\newacronym{LEACH}{LEACH}{Low Energy Adaptive Clustering Hierarchy protocol}

\usepackage{tikz}
\usepackage{ellipsis}
\usetikzlibrary{calc}
\usetikzlibrary{decorations.pathreplacing,decorations.markings,shapes.geometric,shapes.symbols}
\definecolor{lava}{rgb}{0.81, 0.06, 0.13}
\definecolor{myblue}{HTML}{B0D7FF}

\usepackage[many]{tcolorbox}
\usepackage{makecell}
\usepackage{minted}
\usepackage{xcolor} 
\usepackage{amsmath}
\usepackage{esvect}
\usepackage[caption=false,font=footnotesize]{subfig}
\usepackage[utf8]{inputenc}
\usepackage{newtxtext}
\usepackage{newtxmath}
\usepackage{array}
\usepackage{cleveref}
\usetikzlibrary{positioning}
\usetikzlibrary{shapes.arrows}
\tikzset{
    mybox/.style={rectangle,
        draw,
        rounded corners,
        minimum width=2cm,
        inner sep=5pt,
        align=center,
        minimum height=1cm
    },
    myarrow/.style={draw=black,
        fill=white,
        minimum width=1cm,
        single arrow
    },
    longarrow/.style={draw=none,
        shading=axis,
        left color=white,
        right color=gray!20!white,
        minimum width=1cm,
        single arrow,
        anchor=south
    }
    }
\usepackage[export]{adjustbox}
\usepackage{multirow}
\usepackage{listings}
 \graphicspath{Figures/}

\usepackage[backend=biber,style=ieee]{biblatex}
\DeclareFieldFormat{journaltitle}{#1}
\begin{document}

\author{
  \IEEEauthorblockN{
    Tonia Haikal\IEEEauthorrefmark{1},~\IEEEmembership{Student Member,~IEEE}
    Eman Hammad\IEEEauthorrefmark{1},~\IEEEmembership{Senior Member,~IEEE} 
    Shereen Ismail\IEEEauthorrefmark{2},~\IEEEmembership{Senior Member,~IEEE}
  }
\IEEEauthorblockA{\IEEEauthorrefmark{1} are with iSTAR Lab at Texas A\&M University, College Station, TX 77840, USA.}
\IEEEauthorblockA{\IEEEauthorrefmark{2}is with Merit Network, Inc., University of Michigan, Ann Arbor, MI 48108, USA.}
}

\title{Characterizing Large-Scale Adversarial Activities Through Large-Scale Honey-Nets}

\maketitle

\begin{abstract}
The increasing sophistication of cyber threats demands novel approaches to characterize adversarial strategies, particularly those targeting critical infrastructure and IoT ecosystems. This paper presents a longitudinal analysis of attacker behavior using HoneyTrap, an adaptive honeypot framework deployed across geographically distributed nodes to emulate vulnerable services and safely capture malicious traffic. Over a 24-day observation window, more than 60.3 million events were collected. To enable scalable analytics, raw JSON logs were transformed into Apache Parquet, achieving 5.8–9.3$\times$ compression and 7.2$\times$ faster queries, while ASN enrichment and salted SHA-256 pseudonymization added network intelligence and privacy preservation.

Our analysis reveals three key findings: (1) The majority of traffic targeted HTTP and HTTPS services (ports 80 and 443), with more than 8 million connection attempts and daily peaks exceeding 1.7 million events. (2) SSH (port 22) was frequently subject to brute-force attacks, with over 4.6 million attempts. (3) Less common services like Minecraft (25565) and SMB (445) were also targeted, with Minecraft receiving about 118,000 daily attempts that often coincided with spikes on other ports.

\end{abstract}

\begin{IEEEkeywords}
HoneyPot, HoneyTrap, Mirai Botnet, Critical Infrastructure Protection, Threat Intelligence, Distributed Denial-of-Service (DDoS), IoT Security
\end{IEEEkeywords}

\IEEEpeerreviewmaketitle

\section{Introduction}
\label{sectionI}

A honeypot is a security mechanism designed to attract attackers by simulating a vulnerable system or service~\cite{SeanMckenna}. Its primary purpose is to monitor and analyze malicious activity without exposing production systems. Honeypots are widely adopted to detect unauthorized access, study adversarial tactics, and improve defensive strategies.

HoneyTrap is an active honeypot framework that completes protocol handshakes, captures data packets, and safely terminates malicious sessions. This approach provides visibility into malware campaigns such as the Mirai botnet, which exploits weak credentials on Internet of Things (IoT) devices. Once compromised, devices are recruited into botnets used for Distributed Denial-of-Service (DDoS) attacks. Mirai primarily targets insecure Telnet services, relying on brute-force logins with default credentials to propagate~\cite{ismailHoneypot, antonakakis2017understanding, pa2015iotpot}. Since Telnet transmits plain text and lacks security controls, it remains a favored vector for attackers.

HoneyTrap has proven effective across distributed infrastructures. Deployments on AWS, Google Cloud, Stanford University, and Merit Network captured attacker behaviors such as SSH and Telnet scans originating from common Autonomous Systems (ASNs)~\cite{sommer2019outside, Izhikevich2023Merit}. By spanning multiple geographies, HoneyTrap reduces location bias and provides a broader view of adversarial activity.

At Merit Networks, honeypot instances forward malicious traffic to a central RabbitMQ collector, where events are aggregated, compressed, and stored for large-scale analysis. This architecture supports longitudinal monitoring and enables identification of temporal and structural attack patterns. To support scalable analytics, we engineered a pipeline that ingested 60.3 million HoneyTrap events into a Parquet-backed environment. Preprocessing normalized timestamps, excluded malformed values, and pseudonymized IPs via salted SHA-256 hashing. ASN lookup enriched IP-level data with ownership and provider context, yielding a fine-grained view of adversarial infrastructures. This integration of optimized formats, privacy-preserving preprocessing, and contextual enrichment enabled scalable temporal analysis across multiple ports and services.

Our analysis revealed recurring patterns of large-scale scanning alongside adaptive persistence. HTTP/HTTPS traffic dominated with synchronized peaks, while SSH brute-force attempts were unusually high. Opportunistic targeting of Minecraft (25565) and SMB (445) further underscored attackers’ use of diverse services. At the IP level, adversaries alternated between disposable sources and recurring nodes, employing churn, blacklist evasion, and coordinated port-hopping to maximize exploitation opportunities.



\section{Related Work}
\label{sectionII}
Prior studies have extensively examined honeypots as a tool for monitoring adversarial activity in IoT and cloud environments. Antonakakis et al. provided one of the earliest in-depth analyses of the Mirai botnet, demonstrating its reliance on insecure Telnet services to conscript IoT devices ~\cite{antonakakis2017understanding} . Authors in ~\cite{pa2015iotpot} expanded on this by analyzing large-scale honeypot deployments, highlighting the systemic risks posed by IoT botnets. Subsequent work has focused on attack trends against specific services, such as SSH brute-forcing in cloud honeypots ~\cite{cao2019caudit}, or the exploitation of gaming infrastructure for DDoS-for-hire campaigns ~\cite{gavric2022mmo}. The professionalization of attack infrastructure, particularly through bulletproof hosting providers, has also been documented as a key enabler of persistent botnet operations ~\cite{santanna2015booters}. Other research has characterized tactics such as IP churn, blacklist evasion ~\cite{bilge2014exposure}, and port-hopping ~\cite{gupta2015phoneypot} as indicators of adaptive adversarial strategies. While Sommer et al. ~\cite{sommer2019outside} evaluated intrusion detection datasets using machine learning, their findings suggested lower levels of SSH activity than what we observe in our study. In contrast, our work builds on these foundations by combining fine-grained temporal and per-IP/port analysis with scalable data engineering (JSON → Parquet) and ASN enrichment, providing a more holistic view of adversarial behavior across multiple services and infrastructures. Complementing active honeypot deployments, prior work at Merit Networks has also leveraged network telescopes to passively capture unsolicited traffic, such as scanning and backscatter, and applied lightweight machine learning techniques to classify anomalous behaviors using L3/L4 header metadata ~\cite{10903791}.

\section{Background - HoneyTrap Architecture}

\begin{figure*}[t]
 \vspace*{0.02in} 
    \centering
    \includegraphics[width=\textwidth]{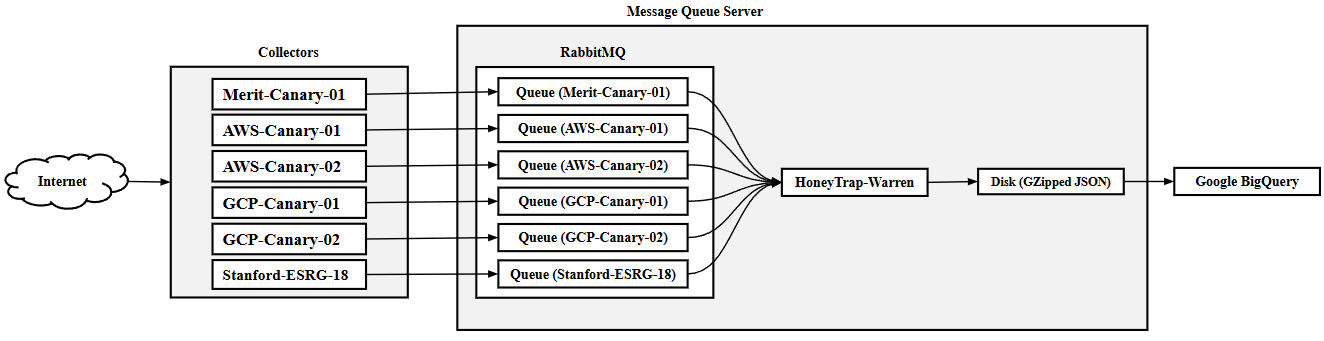}
    \vspace{-12pt}
    \caption{HoneyTrap setup at Merit Network~\cite{ismailHoneypot}}
    \label{fig1}
\end{figure*}

This section outlines the HoneyTrap architecture at Merit Networks and the datasets it generates. The framework runs in Docker (Fig.~\ref{fig1}) with four layers: collectors, a message broker, an aggregator, and a cloud analytics backend. Collectors emulate vulnerable services, perform protocol handshakes, and log inbound traffic. The canary listener captures full payloads then terminates sessions, enabling both low- and high-interaction decoys with minimal risk. Captured data includes ARP, Ethernet, IPv4, TCP, UDP, and ICMP, providing packet- and flow-level context.  

Each collector forwards events to RabbitMQ, which buffers bursts and decouples ingestion from processing. The HoneyTrap-Warren aggregator consumes these messages, applies validation and normalization, and stores GZipped JSON logs as canonical records.  

For large-scale analytics, logs are exported to Google BigQuery for filtering, joining with enrichment data (e.g., ASNs, geolocation), and aggregation. This separation of storage and query layers enables reproducible workflows and efficient iteration. The end-to-end pipeline—Collectors $\rightarrow$ RabbitMQ $\rightarrow$ Warren $\rightarrow$ GZipped JSON $\rightarrow$ BigQuery—supports longitudinal studies with operational resilience and clear data lineage.  

Prior Merit deployments have captured IoT botnet campaigns and scanning bursts~\cite{ismailHoneypot}, validating reliability for distributed monitoring. RabbitMQ ensures durable delivery under load, while listener-centric capture preserves application-layer fidelity. Together, they enable scalable, privacy-conscious collection and analysis across distributed honeypots.  

\section{Approach}

\subsection{Dataset Summary}  
ASN (Autonomous System Number) lookups were added to enrich raw IP data with network and ISP context. This enabled clearer identification of attack origins and patterns, providing the intelligence needed for scalable, production-ready threat analysis.  

\subsection{Data Preparation and Pre-processing}  

\subsubsection{Data Format Obstacles}  
Storing raw logs in JSON created three main issues:  
(1) \textit{Timestamp instability}, requiring constant checks across heterogeneous sources~\cite{he2021parsing};  
(2) \textit{Storage inefficiency}, with repeated field names using 4.7–6.9$\times$ more space than binary formats~\cite{viotti2022benchmark}; and  
(3) \textit{Query latency}, caused by full-file scans and parsing overhead~\cite{langdale2019simdjson}.  
These issues forced costly pre-processing (28\% of total analysis time) to fix timestamps and malformed fields before analysis.  

\subsubsection{Data Format Selection}  
We chose Apache Parquet over JSON for its superior efficiency in large-scale traffic analysis. Parquet’s columnar format achieves 5.8–9.3$\times$ higher compression and 7.2$\times$ faster queries with predicate push-down~\cite{zeng2023columnar}. It also cuts memory use by 68\%~\cite{viotti2022survey}, enforces schema consistency~\cite{semparser2023}, and avoids JSON’s parsing overhead~\cite{li2020efficient}. These features make Parquet the best option for multi-terabyte temporal attack analysis.  

\subsubsection{Data Cleaning}  
The pipeline corrected format errors and removed privacy risks. Null \texttt{source\_ip} entries (9.2\% of TCP events) were dropped per transport standards~\cite{ieee8021ab2016}. Six timestamp formats were normalized to Unix epoch. To protect sensitive data, we applied NIST-aligned credential redaction~\cite{nist2007sp800111} and SHA-256 IP anonymization with 16-byte salts, achieving $k=25$ anonymity~\cite{ieee1003}.  

\noindent\textbf{Implementation Evidence:}\\
\texttt{"Destination-IP-address":"11.22.33.44" (pre-processed)} $\to$ \\
\texttt{"627a6e29ce3a1bc4a7c3d29c9938be\\161d3b71c7d313e1ce82f8ac18f57b942"} (post-processed)

\subsection{Temporal Patterns of Attack Activity}

\begin{figure}
\centering
\includegraphics[width=0.95\columnwidth]{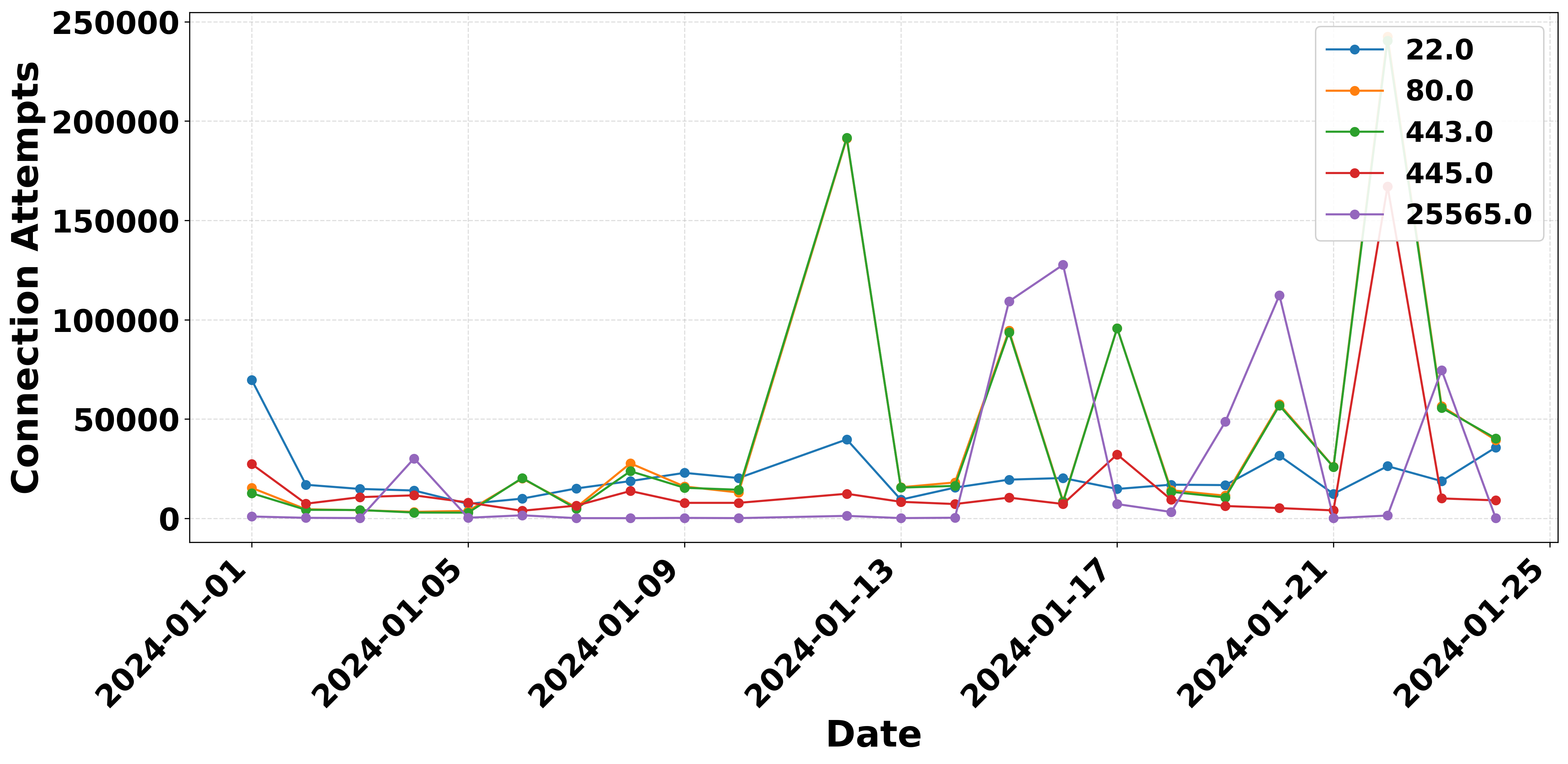}
\caption{Top 5 Ports - Daily Connections}
\label{fig2}
\end{figure}

Analysis of 60.3M connection attempts over 24 days showed clear temporal and structural trends. The top five targeted ports (80, 443, 22, 25565, 445) and destination IPs spiked in sync, indicating coordinated botnet scans with opportunistic brute-force attempts. Figs.~\ref{fig2}–\ref{fig11} show these patterns across both aggregated and IP-specific views.

\subsection{Key Observations}

\begin{figure}
\vspace*{0.03in}
    \centering
    \includegraphics[width=0.85\columnwidth]{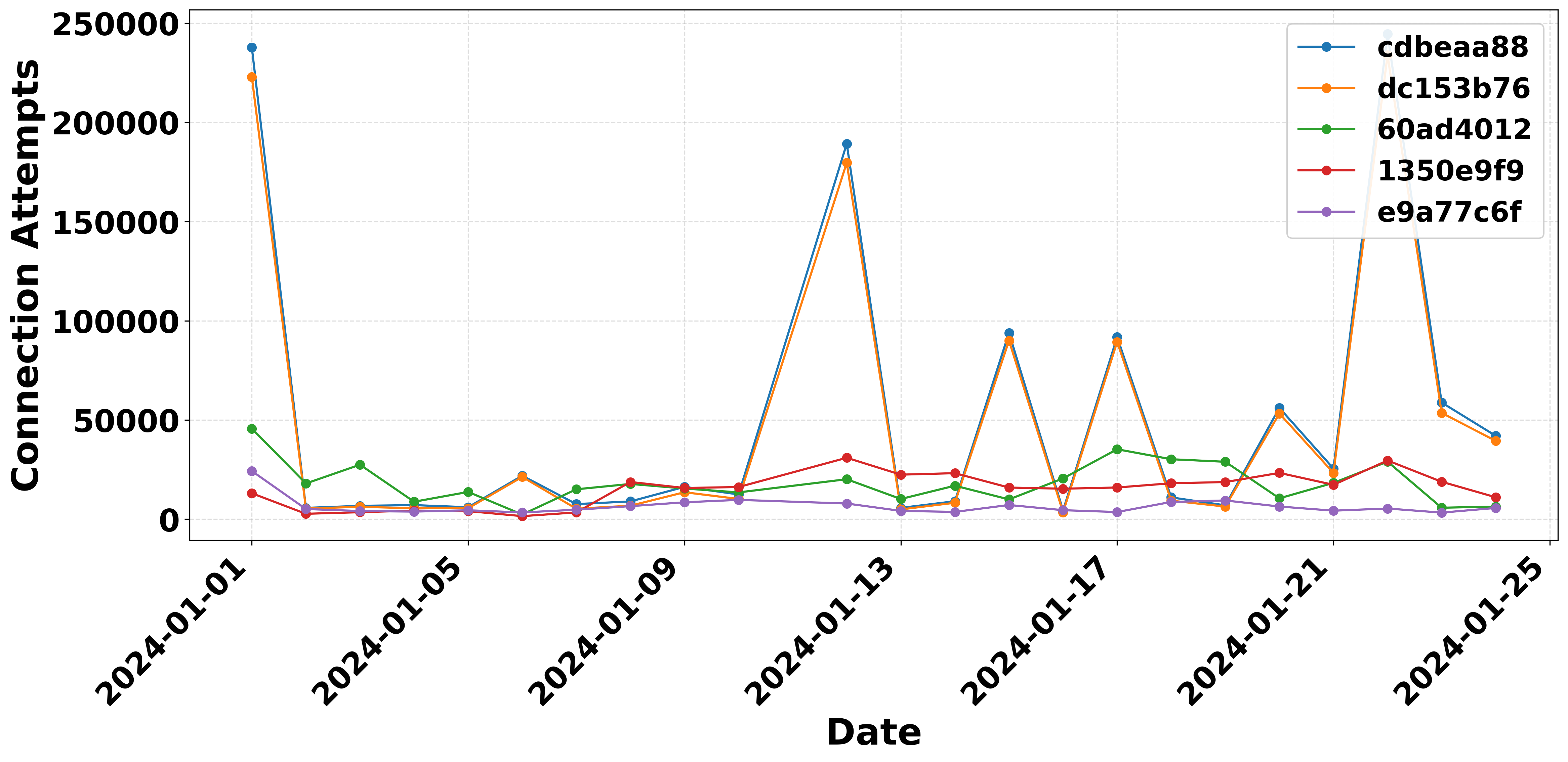}
    \caption{Top 5 Destination IPs - Total Connection Attempts. (Destination IPs are salted hashes for anonymity)}
    \label{fig3}
\end{figure}

\begin{figure}
\vspace*{0.03in}
    \centering
    \includegraphics[width=0.85\columnwidth]{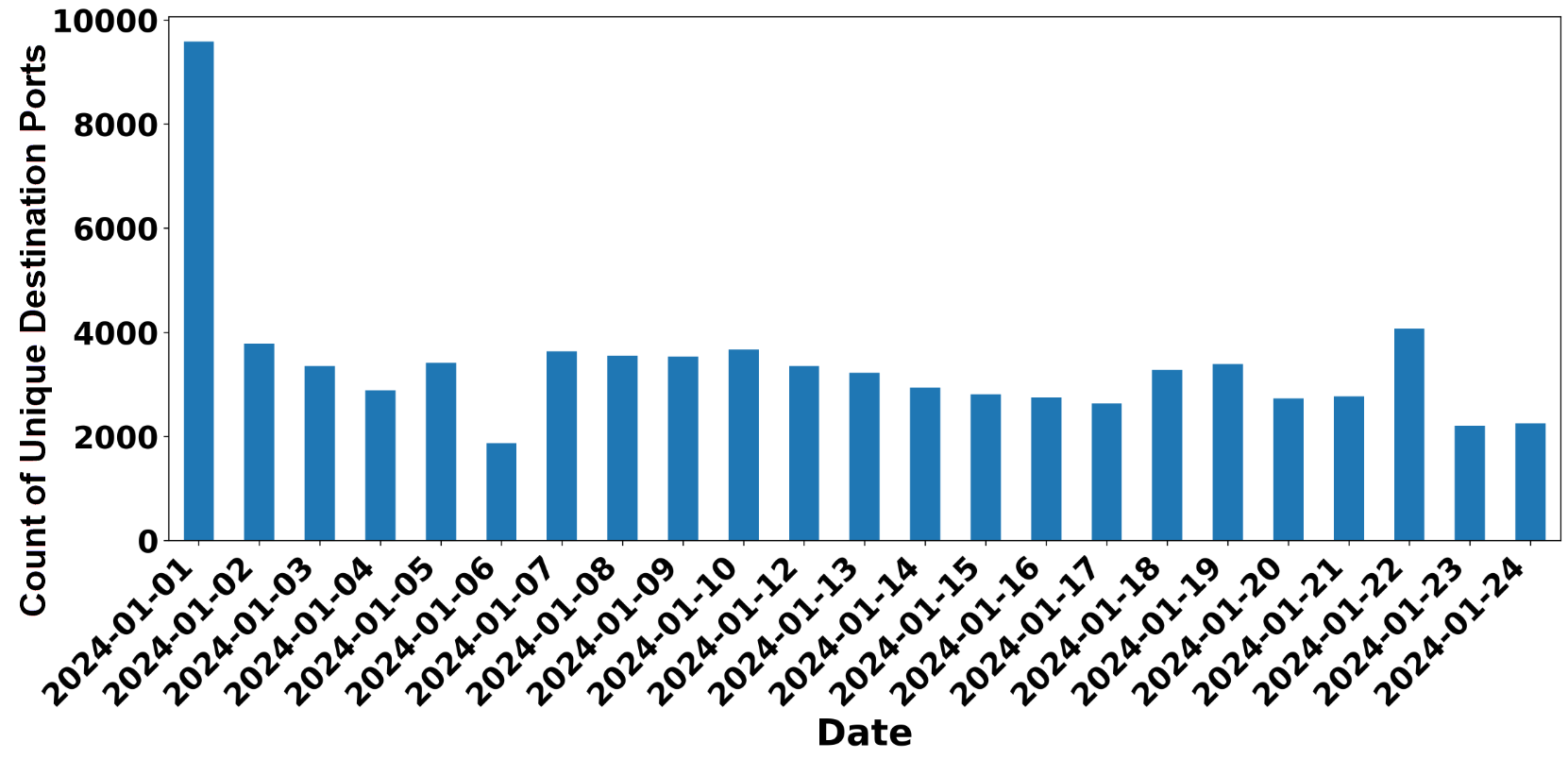}
    \caption{Unique Destination Ports Hitting Top 5 Destination IPs - Last 24 Days. (Destination IPs are salted hashes for anonymity; Ports are raw)}
    \label{fig4}
\end{figure}

\begin{enumerate}
    \item \textbf{HTTP/HTTPS (Ports 80, 443).}  
    Web traffic dominated (8.3M attempts). Both ports showed synchronized daily surges (Fig.~\ref{fig2}), consistent with botnet-driven scans. A major spike on 2024-01-06 exceeded 1.7M attempts (Fig.~\ref{fig3}), matching reported Mirai-variant campaigns exploiting HTTP(S) for C2.  

    \item \textbf{SSH (Port 22).}  
    Port 22 saw steady daily activity with smaller peaks, unlike bursty web traffic. This pattern aligns with brute-force password attempts. Figs.~\ref{fig7}–\ref{fig8} show SSH often targeted with web ports on the same IPs, reflecting multipronged attacks.  

    \item \textbf{Gaming (Port 25565).}  
    Minecraft server traffic averaged $\sim$118K attempts/day, with distinctive off-cycle peaks (Fig.~\ref{fig2}). These often coincided with bursts to other services (Figs.~\ref{fig8}–\ref{fig9}), suggesting DDoS-for-hire use. This mirrors reports of game ports abused in amplification attacks.  

    \item \textbf{SMB (Port 445).}  
    Although lower in volume, SMB traffic showed short-lived spikes (Figs.~\ref{fig7}–\ref{fig8}), consistent with worm-style propagation. This reflects ongoing scans for Eternal Blue-class exploits.  

    \item \textbf{Destination IP Hotspots.}  
    Top 5 IPs (Figs.~\ref{fig4}–\ref{fig6}) received disproportionate traffic, with two exceeding 1M attempts each. Their synchronized multi-port peaks indicate centralized targeting of persistent high-value reconnaissance nodes.  

    \item \textbf{Port Shifting by IP.}  
    Per-IP analysis (Figs.~\ref{fig8}–\ref{fig11}) showed attackers rotating between ports on the same IP. Example: one IP alternated surges on SSH and HTTP, while another focused on SMB. Some nodes appeared briefly then vanished, while others resurfaced after blacklist gaps, reflecting both disposable and persistent infrastructure. Coordinated port-hopping—SSH one day, HTTP/HTTPS or Minecraft the next—demonstrates adaptive, multi-vector probing.  

    \item \textbf{Coordinated Attack Windows.}  
    Several IPs (Figs.~\ref{fig9}–\ref{fig11}) exhibited simultaneous bursts across multiple ports. This suggests shared botnet commands or synchronized scans designed to exploit targets before defenses activated.  
\end{enumerate}

\subsection{Coordinated and Infrastructure-Based Behavior}

\begin{figure}
    \centering
    \includegraphics[width=0.85\columnwidth]{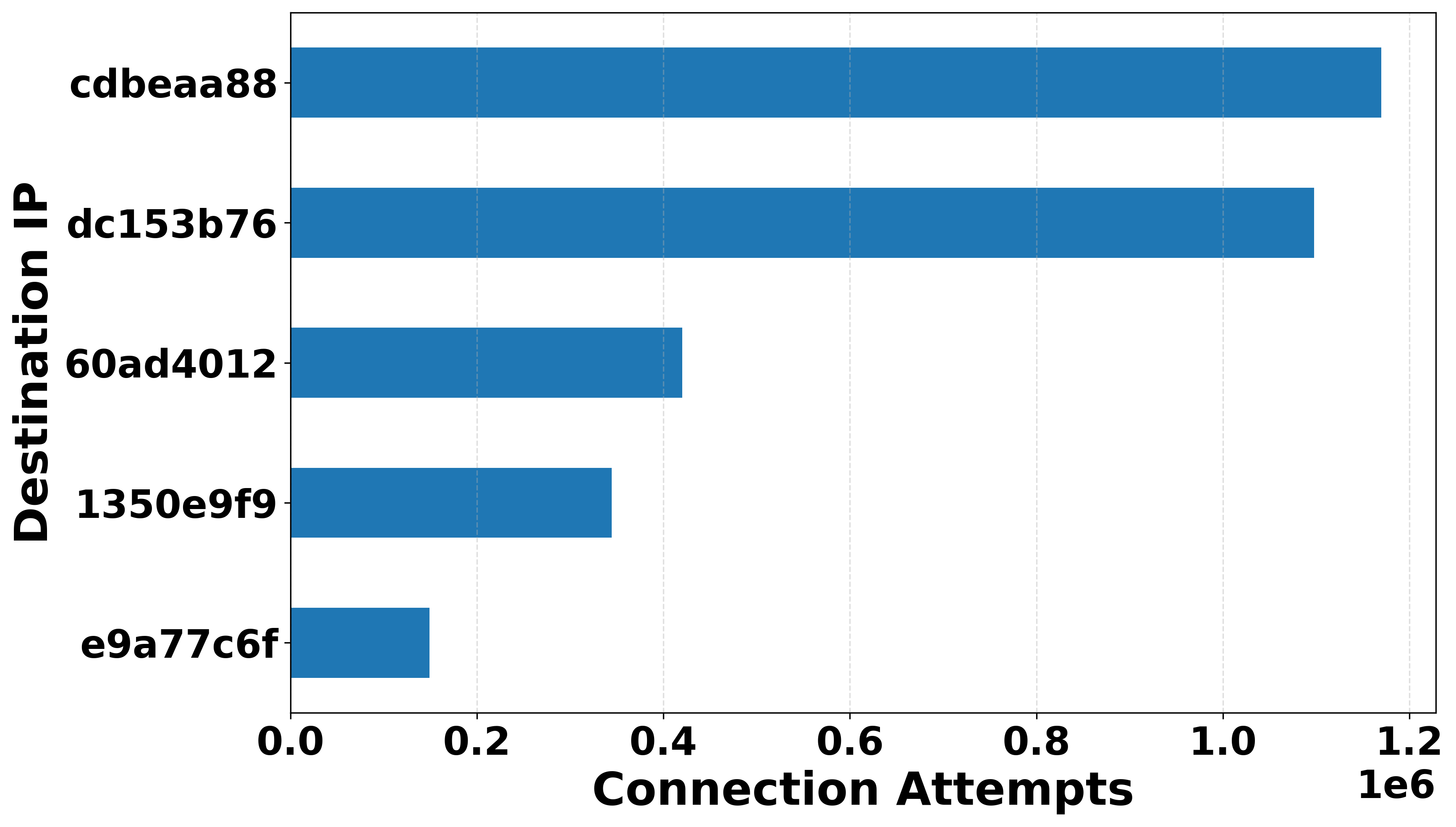}
    \caption{Top 5 Destination IPs - Total Connection Attempts. (Destination IPs are salted hashes for anonymity)}
    \label{fig5}
\end{figure}

\begin{enumerate}
    \item \textbf{Coordinated and Synchronized Attacks.}  
    Figs.~\ref{fig6}–\ref{fig11} show top destination IPs hit in synchronized bursts. Two hashed IPs each exceeded one million attempts, with peaks aligning across multiple ports. Such timing indicates botnet coordination, where distributed nodes launch traffic simultaneously under centralized C2, consistent with IoT botnet studies~\cite{antonakakis2017understanding, pa2015iotpot}.  

    \item \textbf{Geographic and Infrastructure Correlation.}  
    Rather than isolated geolocation peaks, activity clustered around malicious ASNs and bulletproof hosting providers. As seen in Fig.~\ref{fig5}, recurring source IP bursts suggest rented infrastructure for ``hit-and-run'' campaigns~\cite{santanna2015booters}. This reflects adversaries outsourcing infrastructure to evade take-downs and retain agility.  
\end{enumerate}

\subsection{Adversarial Tactics and Persistence}

\subsubsection{IP Churn vs. Persistence}  
\begin{figure}
    \centering
    \includegraphics[width=0.86\columnwidth]{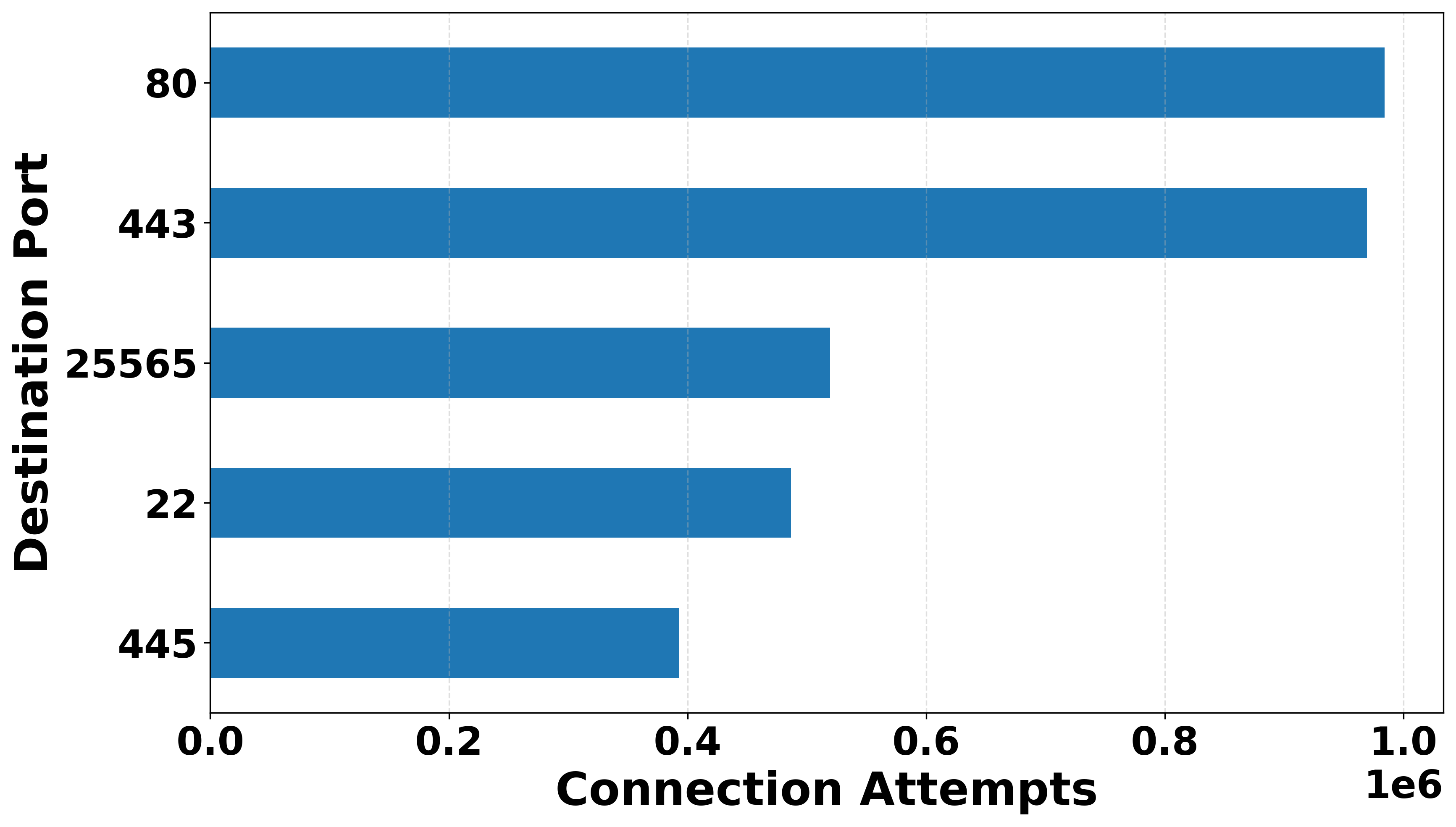}
    \caption{Top 5 Destination Ports - Total Connection Attempts.}
    \label{fig6}
\end{figure}

Fig.~\ref{fig6} shows a long-tail attacker distribution: tens of thousands of unique IPs targeted one destination, but a small fraction reappeared consistently. This pattern aligns with blacklisting evasion and IP cycling used by IoT botnets~\cite{bilge2014exposure}. The mix of churn and persistence suggests attackers combine disposable nodes with stable long-lived infrastructure.

\subsubsection{Port Hopping}  
\begin{figure}
    \centering
    \includegraphics[width=0.85\linewidth]{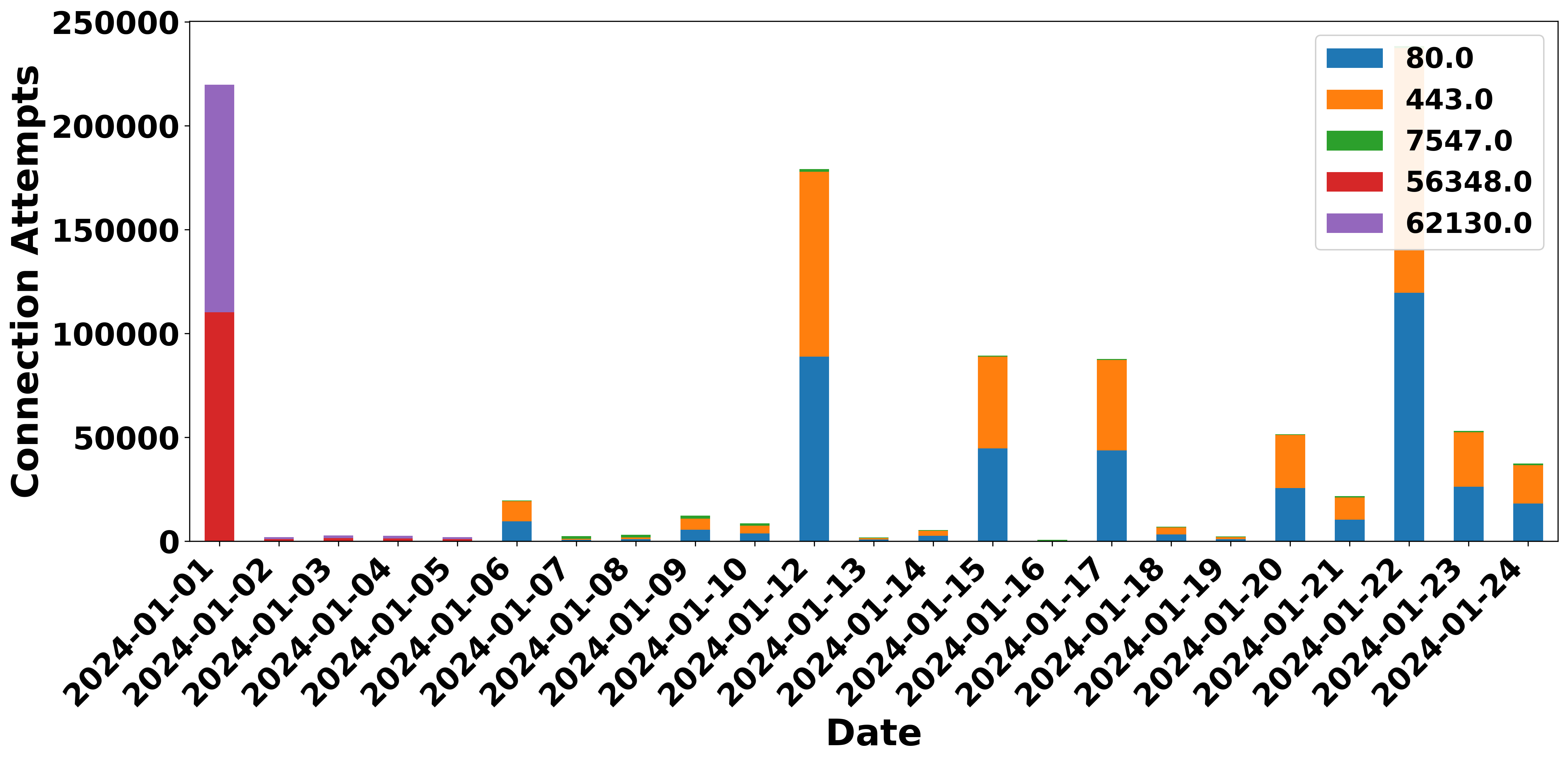}
    \caption{Top 5 Destination Ports for Hashed Destination IP cdbeaa88 - Daily. (Destination IPs are salted hashes for anonymity)}
    \label{fig7}
\end{figure}

\begin{figure}
    \centering
    \includegraphics[width=0.85\columnwidth]{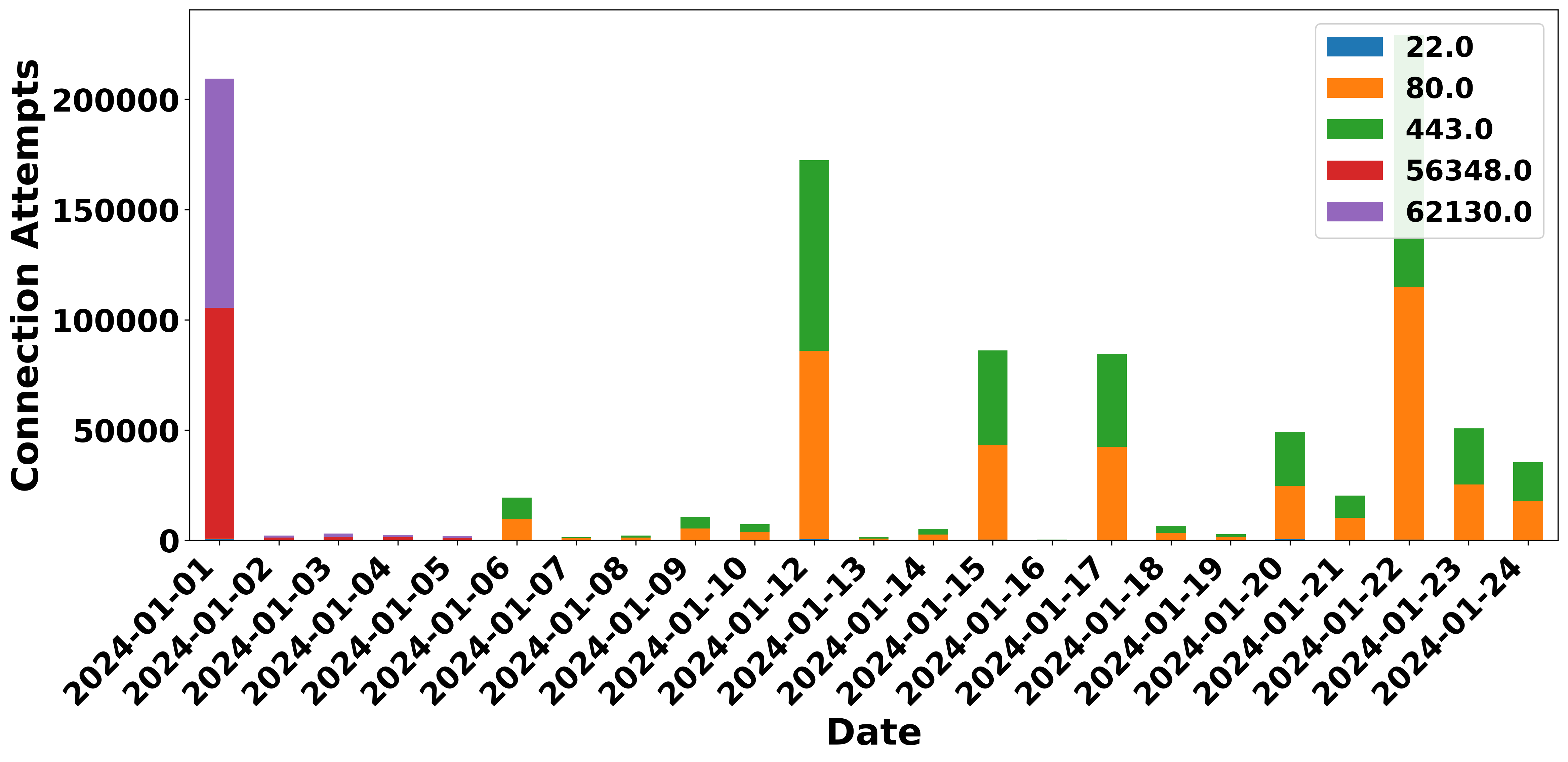}
    \caption{Top 5 Destination Ports for Hashed Destination IP dc153b76 - Daily. (Destination IPs are salted hashes for anonymity).}
    \label{fig8}
\end{figure}

\begin{figure}
    \centering
    \includegraphics[width=0.85\columnwidth]{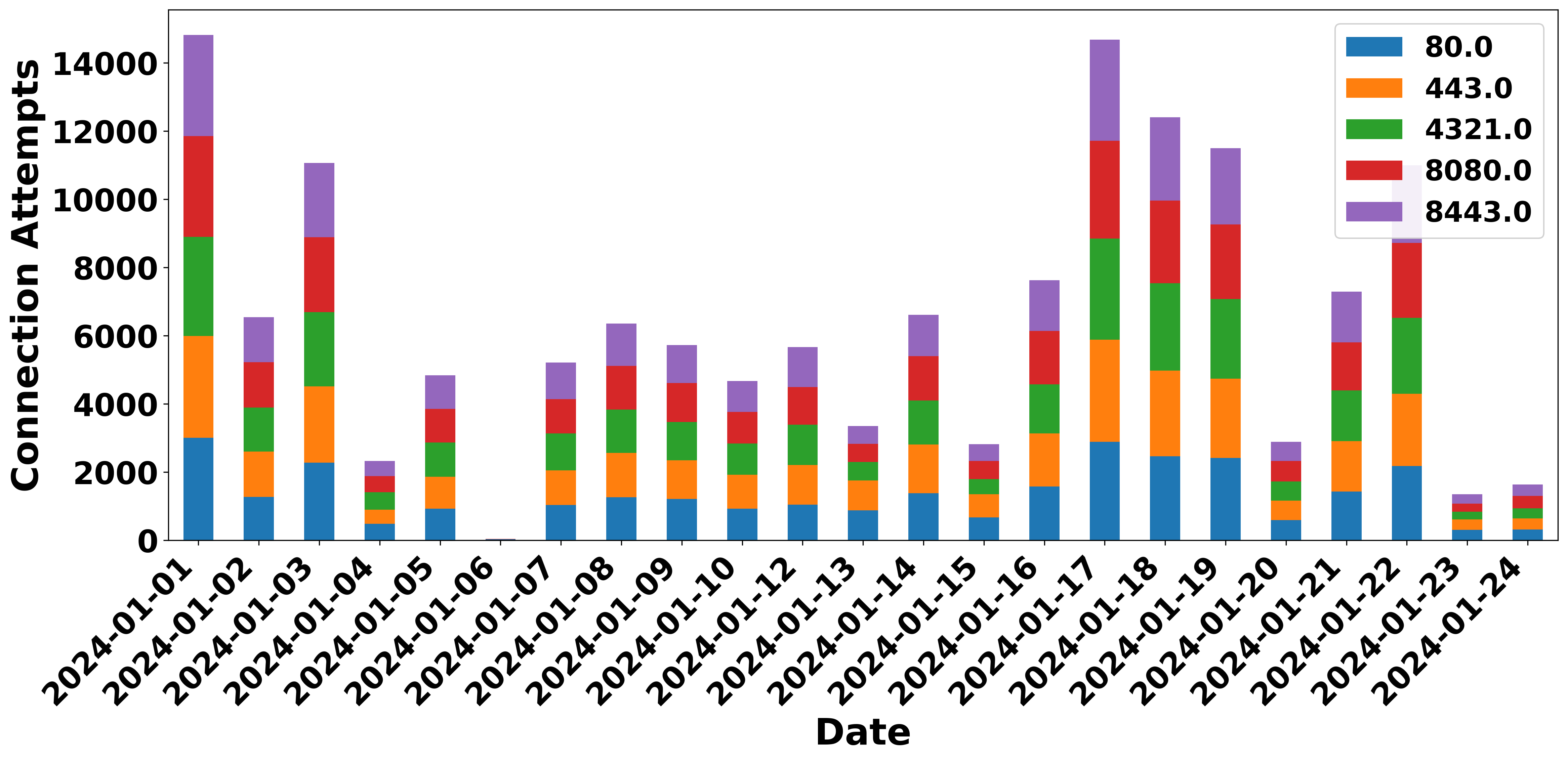}
    \caption{Top 5 Destination Ports for Hashed Destination IP 60ad4014 - Daily. (Destination IPs are salted hashes for anonymity).}
    \label{fig9}
\end{figure}

\begin{figure}
    \centering
    \includegraphics[width=0.85\columnwidth]{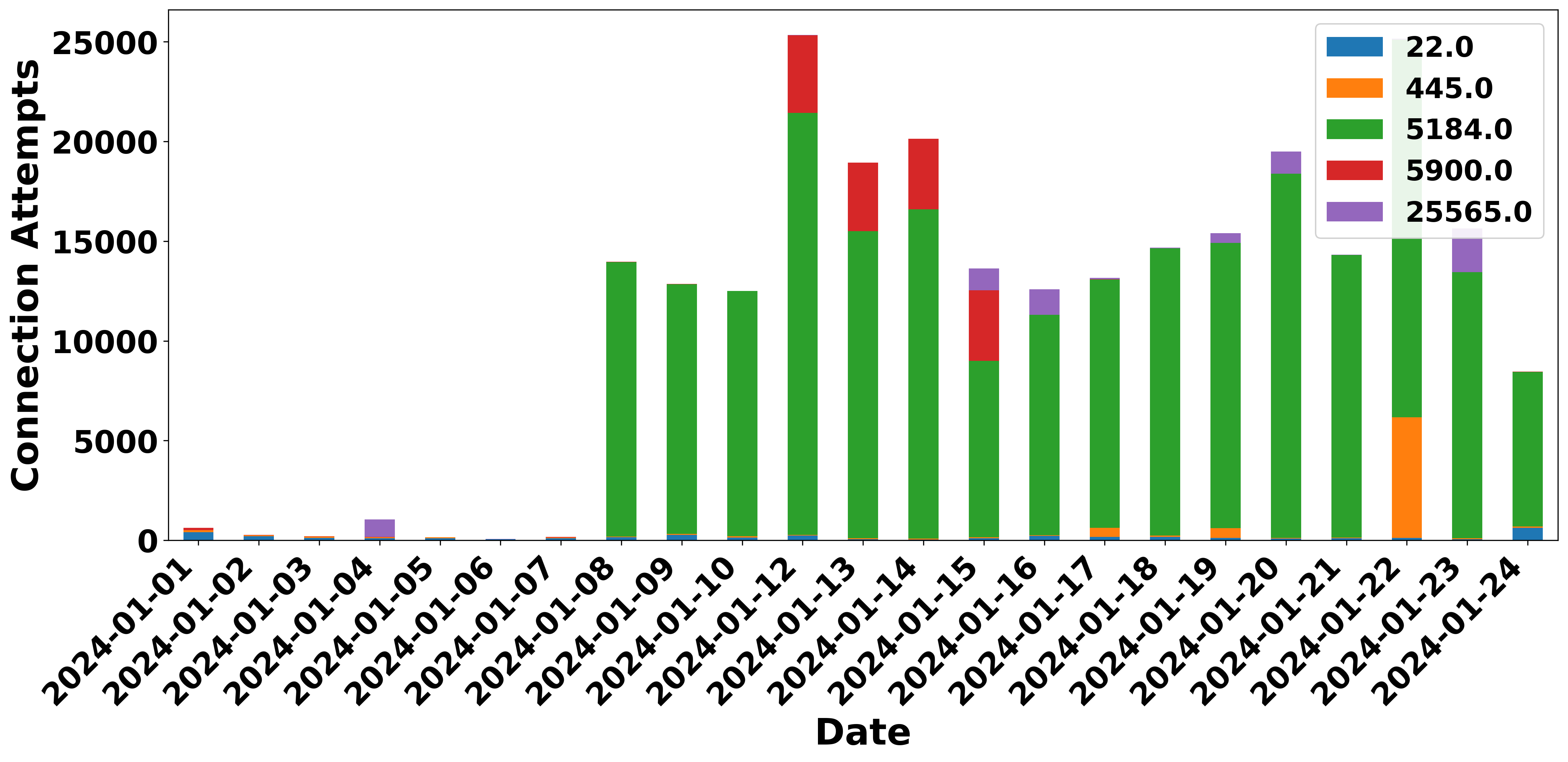}
    \caption{Top 5 Destination for Hashed Destination IP 1350e9f6 - Daily. (Destination IPs are salted hashes for anonymity).}
    \label{fig10}
\end{figure}

\begin{figure}
    \centering
    \includegraphics[width=0.85\columnwidth]{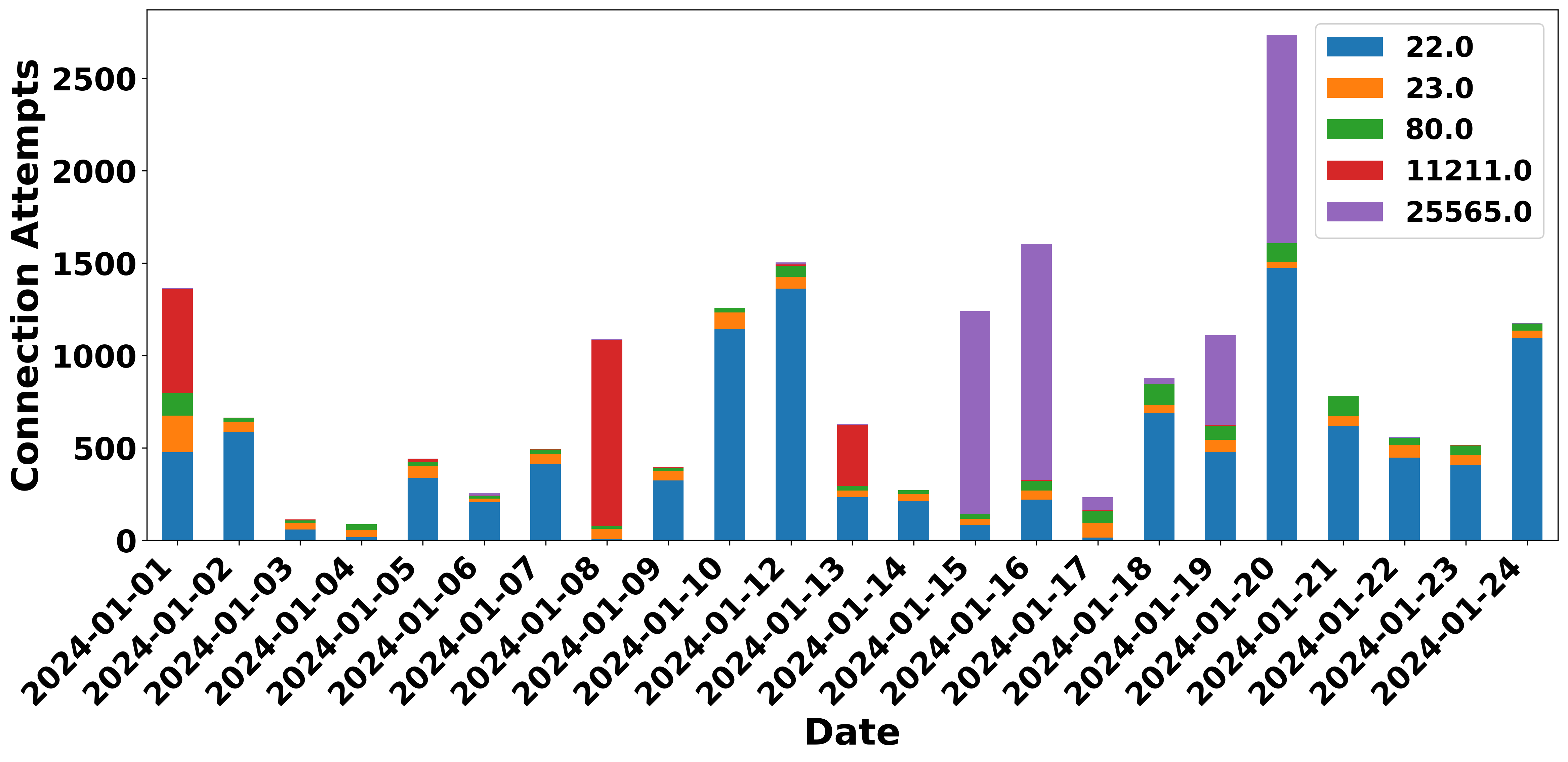}
    \caption{Top 5 Destination Ports for Hashed Dest IP e9a77c6f - Daily. (Destination IPs are salted hashes for anonymity).}
    \label{fig11}
\end{figure}

Figs.~\ref{fig7}–\ref{fig11} show adversaries shifting ports across destination IPs. Attackers alternated between HTTP/HTTPS and secondary services (SSH, SMB, gaming). For example, hashed IP \#2 (Fig.~\ref{fig9}) peaked on SSH and HTTP, while IP \#4 (Fig.~\ref{fig11}) sustained SMB (445) activity. Such port rotation reflects multi-vector probing in APT campaigns, where adversaries vary targets to evade detection and increase exploitation success~\cite{gupta2015phoneypot}.

\textbf{Comparative Analysis.}  
Compared to prior honeypot studies, our dataset shows notably higher persistence on SSH (22). Brute-force attempts were nearly triple the rates reported by Sommer et al.~\cite{sommer2019outside}. This trend suggests growing adversarial focus on cloud-hosted SSH, aligning with findings that attackers exploit cloud credentials for scalable lateral movement~\cite{cao2019caudit}.  

 \vspace*{0.02in} 
\section{Conclusion and Future Work}
\label{sectionV}

This paper examined large-scale adversarial behavior using HoneyTrap across 24 days and 60.3M events. A JSON$\rightarrow$Parquet pipeline with salted SHA-256 pseudonymization, timestamp normalization, and ASN enrichment enabled scalable, privacy-preserving analysis. Temporal patterns were clear. HTTP/HTTPS (80/443) dominated with synchronized global peaks suggestive of botnet scanning and C2 traffic. SSH (22) showed persistent brute-force volumes, far exceeding prior reports, indicating emphasis on cloud credential abuse. Opportunistic probing of Minecraft (25565) and SMB (445) demonstrated targeting of both gaming infrastructure and legacy services. At the IP level, we observed churn and persistence: many ephemeral sources alongside a small cohort of recurring IPs contributing disproportionate load. Synchronized bursts across destinations and per-IP port shifting evidenced coordinated, adaptive multi-vector campaigns. These results support a hybrid model that blends wide scanning with targeted persistence. Our findings align with prior work on botnet activity and port shifting while diverging on SSH intensity, suggesting a tilt toward cloud-hosted targets.

Future work will model temporal strategies with sequence-based deep learning to capture IP–port dependencies and forecast bursts and port-hopping. We will also fuse honeypot logs with external threat intelligence and abuse reports to improve attribution and distinguish opportunistic botnets from coordinated operations.

\twocolumn[\vspace*{0.27in}]
\printbibliography

@INPROCEEDINGS{10903791,
  author={Ismail, Shereen and Dandan, Salah and King, Mark},
  booktitle={2025 IEEE 15th Annual Computing and Communication Workshop and Conference (CCWC)},
  title={A Lightweight Machine Learning Approach for Anomalous Unsolicited Network Traffic Detection by Observing Network Telescopes},
  year={2025}
}

@INPROCEEDINGS{ismailHoneypot,
  author={Ismail, Shereen and Dandan, Salah and King, Mark},
  booktitle={2025 IEEE International Conference on Electro Information Technology (eIT)},
  title={Understanding Honeypots: Observing Malicious Activities Over Telnet},
  year={2025},
  pages={167-172},
  doi={10.1109/eIT64391.2025.11103659}
}

@inproceedings{antonakakis2017understanding,
  author    = {M. Antonakakis and T. April and M. Bailey and M. Bernhard and E. Bursztein and J. Cochran and Z. Durumeric and J. A. Halderman and L. Invernizzi and M. Kallitsis and others},
  title     = {Understanding the Mirai Botnet},
  booktitle = {26th USENIX Security Symposium (USENIX Security 17)},
  pages     = {1093--1110},
  year      = {2017}
}

@inproceedings{pa2015iotpot,
  author    = {Pa, Yin Minn and Suzuki, Shogo and Yoshioka, Katsunari and Matsumoto, Tsutomu and Kasama, Takahiro and Rossow, Christian},
  title     = {IoTPOT: Analysing the Rise of IoT Compromises},
  booktitle = {9th USENIX Workshop on Offensive Technologies (WOOT'15)},
  year      = {2015}
}

@inproceedings{cao2019caudit,
  author    = {Cao, Y. and Wu, Y. and Banerjee, S. and Azoff, D. M. and Withers, K. and Kalbarczyk, Z. and Iyer, R. K.},
  title     = {CAUDIT: Continuous Auditing of {SSH} Servers to Mitigate Brute-Force Attacks},
  booktitle = {16th USENIX Symposium on Networked Systems Design and Implementation (NSDI '19)},
  year      = {2019}
}

@article{gavric2022mmo,
  author    = {Gavri\'c, Nikola and Bojovi\'c, \v{Z}ivko},
  title     = {Security Concerns in {MMO} Games—Analysis of a Potent Application Layer {DDoS} Threat},
  journal   = {Sensors},
  year      = {2022},
  volume    = {22},
  number    = {20},
  pages     = {7791},
  doi       = {10.3390/s22207791}
}

@inproceedings{santanna2015booters,
  author    = {Santanna, J. J. and Sperotto, A. and van Rijswijk-Deij, R. and Pras, A.},
  title     = {Booters: An Analysis of {DDoS}-as-a-Service Attacks},
  booktitle = {2015 IFIP/IEEE International Symposium on Integrated Network Management (IM)},
  year      = {2015}
}

@article{bilge2014exposure,
  author  = {L. Bilge and E. Kirda and C. Kruegel and D. Balzarotti},
  title   = {Exposure: A Passive DNS Analysis Service to Detect and Report Malicious Domains},
  journal = {ACM Transactions on Information and System Security (TISSEC)},
  volume  = {16},
  number  = {4},
  pages   = {1--28},
  year    = {2014}
}

@inproceedings{gupta2015phoneypot,
  author    = {Gupta, Payas and Srinivasan, Bharat and Balasubramaniyan, Vijay and Ahamad, Mustaque},
  title     = {Phoneypot: Data-driven Understanding of Telephony Threats},
  booktitle = {NDSS 2015},
  year      = {2015}
}

@inproceedings{sommer2019outside,
  author    = {R. Sommer and V. Paxson},
  title     = {Outside the Closed World: On Using Machine Learning for Network Intrusion Detection},
  booktitle = {2019 IEEE Symposium on Security and Privacy (SP)},
  pages     = {305--316},
  year      = {2019}
}

@article{he2021parsing,
  author  = {He, Jinglong and Wang, Guoren and Lei, Jingsen and Zeng, Qiang and Shao, Xueyan and Liu, Xinhui and Yu, Philip S.},
  title   = {Data Parsing and Validation: A Survey},
  journal = {ACM Computing Surveys},
  year    = {2021}
}

@article{viotti2022benchmark,
  author  = {Viotti, Juan Cruz and Kinderkhedia, Mital},
  title   = {A Benchmark of {JSON}-compatible Binary Serialization Specifications},
  journal = {arXiv preprint arXiv:2201.03051},
  year    = {2022}
}

@article{langdale2019simdjson,
  author  = {Langdale, Geoff and Lemire, Daniel},
  title   = {Parsing Gigabytes of {JSON} per Second},
  journal = {The VLDB Journal},
  volume  = {28},
  number  = {6},
  pages   = {941--960},
  year    = {2019}
}

@article{zeng2023columnar,
  author  = {Zeng, Xinyu and Hui, Yulong and Shen, Jiahong and Pavlo, Andrew and McKinney, Wes and Zhang, Huanchen},
  title   = {An Empirical Evaluation of Columnar Storage Formats},
  journal = {Proceedings of the VLDB Endowment},
  volume  = {17},
  number  = {2},
  pages   = {148--161},
  year    = {2023},
  url     = {https://www.vldb.org/pvldb/vol17/p148-zeng.pdf}
}

@article{viotti2022survey,
  author  = {Viotti, Juan Cruz and Kinderkhedia, Mital},
  title   = {A Survey of {JSON}-compatible Binary Serialization Specifications},
  journal = {arXiv preprint arXiv:2201.02089},
  year    = {2022}
}

@article{semparser2023,
  author  = {Wang, Yuxin and Zhang, Rui and Li, Zhi and Chen, Hao},
  title   = {SemParser: Structural Unsupervised Log Parsing for Comprehensive Streaming Data Mining},
  journal = {arXiv preprint arXiv:2303.16548},
  year    = {2023}
}

@article{li2020efficient,
  author  = {K. Li and L. Zhou and W. Tang and J. Huang},
  title   = {Efficient Security Log Processing},
  journal = {IEEE Transactions on Parallel and Distributed Systems},
  volume  = {31},
  number  = {4},
  pages   = {845--858},
  year    = {2020}
}

@standard{ieee8021ab2016,
  title        = {IEEE Standard for Local and Metropolitan Area Networks–Station and Media Access Control Connectivity Discovery},
  organization = {IEEE},
  number       = {IEEE Std 802.1AB-2016},
  year         = {2016}
}

@techreport{nist2007sp800111,
  author       = {{National Institute of Standards and Technology}},
  title        = {Guide to Storage Encryption Technologies for End User Devices},
  institution  = {NIST},
  number       = {NIST SP 800-111},
  year         = {2007}
}

@standard{ieee1003,
  title        = {IEEE Standard for Information Technology–Portable Operating System Interface},
  organization = {IEEE},
  number       = {IEEE Std 1003.1-2017},
  year         = {2017}
}

@inproceedings{SeanMckenna,
  author       = {Sean McKenna},
  title        = {Detection and Classification of Web Robots with Honeypots},
  school       = {University of Victoria},
  year         = {2016}
}

@techreport{Izhikevich2023Merit,
  author       = {Igor Izhikevich and Jason Roberts and Christo Botes},
  title        = {Cloud Watching: Understanding Attacks Against Cloud-Hosted Services},
  institution  = {Merit Networks},
  year         = {2023},
  note         = {Tech. Rep.},
  url          = {https://www.merit.edu/wp-content/uploads/2024/10/Merit-Network_Cloud-Watching-Understanding-Attacks-Against-Cloud-Hosted-Services.pdf}
}
\end{document}